\documentclass[aps,pre,reprint,superscriptaddress]{revtex4-2}

\usepackage{bm}
\usepackage{amssymb}
\usepackage{pifont}
\usepackage{amsmath}
\usepackage{subfigure}
\usepackage{graphics}
\usepackage{graphicx}
\usepackage{epsfig}
\usepackage{pstricks}
\usepackage{pst-plot}
\usepackage{pst-slpe}
\usepackage{latexsym}
\usepackage{bm}
\usepackage{url}
\usepackage{pifont}
\usepackage{amsmath}
\usepackage{amsfonts}
\usepackage[margin=2.5cm]{geometry}
\usepackage{float}
\usepackage{makecell}

 \newcommand{\be}{\begin{equation}}
 \newcommand{\ee}{\end{equation}}
 \newcommand{\bea}{\begin{eqnarray}}
 \newcommand{\eea}{\end{eqnarray}}
\usepackage{lineno,hyperref}

\begin{document}


\title{Bond percolation in distorted square and triangular lattices}


\author{Bishnu Bhowmik}
\affiliation{Department of Physics, University of Gour Banga, Malda-732103, India}
\author{Sayantan Mitra}
\affiliation{Department of Physical Sciences, Indian Institute of Science Education and Research Kolkata, Mohanpur, 741246 India}
\author{Robert M. Ziff}
\affiliation{Center for the Study of Complex Systems and Department of Chemical Engineering, University of Michigan, Ann Arbor, Michigan 48109-2800, USA}
\author{Ankur Sensharma}
\affiliation{Department of Physics, University of Gour Banga, Malda-732103, India}
\email[]{itsankur@ugb.ac.in}


\date{\today}

\begin{abstract}
This article presents a Monte Carlo study on bond percolation in distorted square and triangular lattices. The distorted lattices are generated by dislocating the sites from their regular positions. The amount and direction of the dislocations are random, but can be tuned by the distortion parameter $\alpha$. Once the sites are dislocated, the bond lengths $\delta$ between the nearest neighbors change. A bond can only be occupied if its bond length is less than a threshold value called the connection threshold $d$. It is observed that when the connection threshold is greater than the lattice constant (assumed to be $1$), the bond percolation threshold $p_\mathrm{b}$ always increases with distortion. For $d\le 1$, no spanning configuration is found for the square lattice when the lattice is distorted, even very slightly. On the other hand, the triangular lattice not only spans for $d\le 1$, it also shows a decreasing trend for $p_\mathrm{b}$ in the low-$\alpha$ range. These variation patterns have been linked with the average coordination numbers of the distorted lattices. A critical value $d_\mathrm{c}$ for the connection threshold has been defined as the value of $d$ below which no spanning configuration can be found even after occupying all the bonds satisfying the connection criterion $\delta\le d$. The behavior of $d_\mathrm{c}(\alpha)$ is markedly different for the two lattices.     
\end{abstract}


\maketitle


\section{Introduction}
Percolation, introduced by Broadbent and Hammersley in 1957 \cite{Broadbent} is a simple statistical model that exhibits non-trivial critical behavior \cite{Book1}. The model has been applied to a number of problems of physics, specially to understand the   critical behavior of the metal-insulator transition \cite{Ball}, the diamagnetic-ferromagnetic transition \cite{Dotsenko} and others. It has also been applied to study electric \cite{Avella2019, Cheng2020} and magnetic \cite{Yiu2014, Grady2023} properties of solids. Apart from Physics, researchers of various fields have applied this model to explain many natural phenomena like spreading of fires in forests \cite{Albano1, Guisoni}, flow of liquid through a porous media \cite{Hunt}, spreading of epidemic diseases \cite{Moore, Ziff2, Miller, Dipa2023} and many more \cite{Saberi1,Stauffer,Saberi3}.

The mechanism of  two basic variants of percolation, namely, the site percolation and the bond percolation, is as follows. The sites (bonds) can be in two possible states: unoccupied or empty and occupied or filled. Starting from an empty lattice, a site (bond) is selected randomly and is occupied with an occupation probability $p$. Clusters are formed by two or more neighboring occupied sites (bonds). As $p$ increases, clusters get bigger. In percolation, one often searches for end-to-end connectivity, which is called spanning. This happens at a certain value of $p$ called the percolation threshold $p_c (p_\mathrm{b})$. For an infinite lattice, the size of the spanning cluster also becomes infinite. The sudden occurrence of a spanning cluster marks a phase transition. Determination of the percolation threshold and associated critical  exponents to characterize the transition are often the key interests of the physicists. For most systems, the percolation threshold has been found numerically. Analytical results can be obtained for classes of lattices that are related to self-dual hypergraphs \cite{Scullard, Wierman} . Two such analytical results relevant to the present work are the bond percolation thresholds for the square lattice ($p_\mathrm{b}=1/2$) and the triangular lattice ($p_\mathrm{b}=2\sin (\pi /18)$) \cite{Sykes1964}.

Many other variants of percolation have been introduced to suit different natural scenarios. These include site-bond percolation \cite{Tarasevich}, directed percolation \cite{Takeuchi}, bootstrap percolation \cite{Adler}, explosive percolation \cite{Adler,Achlioptas}, correlated percolation \cite{Coniglio} continuum percolation \cite{Hall1985, Mertens} and many others. 

\begin{figure*}
	  \subfigure[]{\includegraphics[scale=0.7]{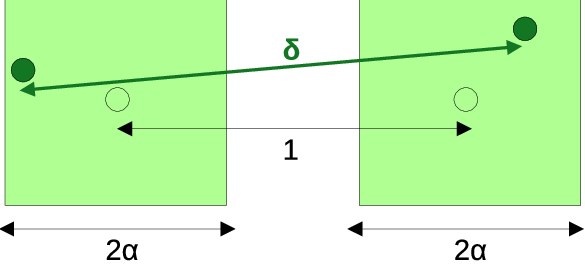}}\hspace{0.5cm}
      \subfigure[]{\includegraphics[scale=0.7]{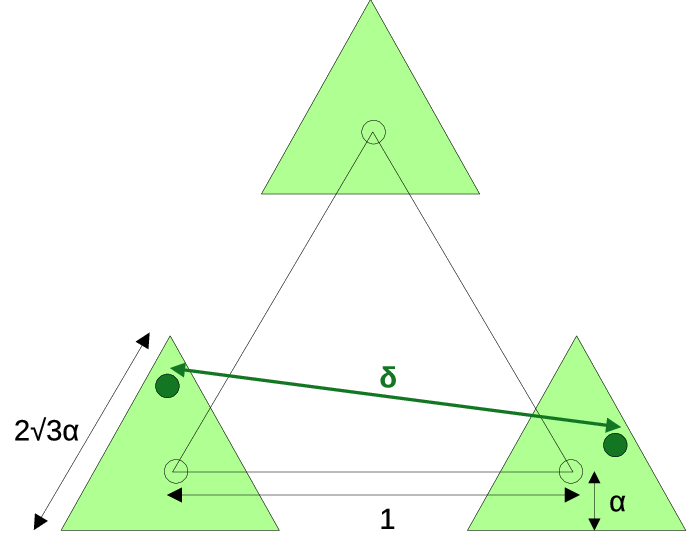}}
\caption{Mechanism of distortion in (a) square and (b) triangular lattices. Green squares and triangles indicate possible regions of the dislocated sites.}
\label{fig:dist_mech}
\end{figure*}

Precise numerical estimates of site and bond percolation thresholds have been calculated for many lattices \cite{Wang,Jacobsen_2015,Manna,Deng,Ballesteros, Gonzalez, Lorentz}. Efforts have also been put behind finding the percolation threshold when extended or complex neighborhood between the sites and bonds are considered \cite{Xun1,Xun2,Xun3,Malarz,Malarz2021,Malarz2022,Malarz2024}. The critical exponents \cite{Liu,Kirkpatrick,Kundu2,Sommers,Hassan} are determined in various models. It turns out that the values of these exponents usually do not depend on the lattice geometry or the type of percolation (site or bond) as long as the dimensionality remains the same \cite{Saberi1, Stauffer,Bollobas_book}. For example, the site and bond percolation thresholds for square and triangular lattices have different values, but the values of the critical exponents are the same for all these four cases.  

Imperfections in a lattice can significantly change its percolation properties. A few articles on percolation in disordered lattices are found in the literature \cite{Kundu1,Centres2010, Spencer}. Previously, we studied site percolation in distorted lattices in two and three dimensions \cite{Sayantan1,Sayantan2,Sayantan3}. The percolation threshold was shown to depend on the interplay between two tunable parameters -- the distortion parameter and the connection threshold. When only nearest neighbor connection is considered and the connection threshold is set at a higher value than the lattice constant, the percolation threshold threshold increases with distortion for both distorted square \cite{Sayantan1} and distorted simple cubic lattices \cite{Sayantan2}. This indicates that spanning becomes difficult with distortion in this regime. When the connection threshold is less than the lattice constant, the square lattice did not span even for a slight distortion. On the other hand, the distorted simple cubic lattice does span in this regime, although the percolation threshold shows a different behavior that includes an initial decline with distortion \cite{Sayantan2}. When higher order connectivity is considered \cite{Sayantan3}, a much more complicated variation pattern has been observed.

In this work, we apply similar ideas to study bond percolation in distorted square and triangular lattices. Since the lattice is distorted, the bond distances between the nearest neighbors change. A bond can only be occupied if its length is below a certain limit, called the connection threshold. Therefore, there is a the lack of freedom in choosing a bond for occupation. It turns out that this restriction has a significant impact on the bond percolation properties of the distorted lattices. The remaining part of the paper is organized as follows. Section \ref{sec:model} is devoted to the distortion mechanism and methodology of the model. The results are presented in Sections \ref{sec:pb} and \ref{sec:dc}. The variations of the percolation threshold in distorted square and triangular lattices are demonstrated in subsections \ref{sec:sqpb} and \ref{sec:tripb} respectively. Precise thresholds are reported in \ref{sec:pbprecise}. In Section \ref{sec:univ}, it is shown that this model belongs to the same universality class of regular percolation. The quantity critical connection threshold has been introduced and its variation with distortion is shown in Section \ref{sec:dc}. We discuss the present results in the perspective of our earlier studies in Section \ref{sec:perspective} and finally summarize in Section \ref{sec:summary}.

\section{The model and the process}\label{sec:model}
Fig.\ref{fig:dist_mech} illustrates the mechanism of how the lattice sites are distorted. Here $\alpha$ is the distortion parameter. Each lattice site of a (i) regular square lattice of unit lattice constant is dislocated to any random point within a square of side $2\alpha$, and (ii) regular triangular lattice of unit lattice constant is dislocated to any random point within an equilateral triangle of side $2\sqrt{3}\alpha$. The new distance between the two dislocated lattice points is denoted by $\delta$. The minimum and maximum values of $\delta$ for square and triangle are 
\begin{eqnarray}\label{eq:minmax}
\delta_\mathrm{min}^\mathrm{Sq}&=&1-2\alpha\\ \nonumber
\delta_\mathrm{Max}^\mathrm{Sq}&=&\sqrt{(1+2\alpha)^2+(2\alpha)^2}\\\nonumber
\delta_\mathrm{min}^\mathrm{Tr}&=&1-2\sqrt{3}\alpha\\\nonumber
\delta_\mathrm{Max}^\mathrm{Tr}&=&1+2\sqrt{3}\alpha \nonumber
\end{eqnarray}

\begin{figure*}
	  \subfigure[]{\includegraphics[scale=0.52]{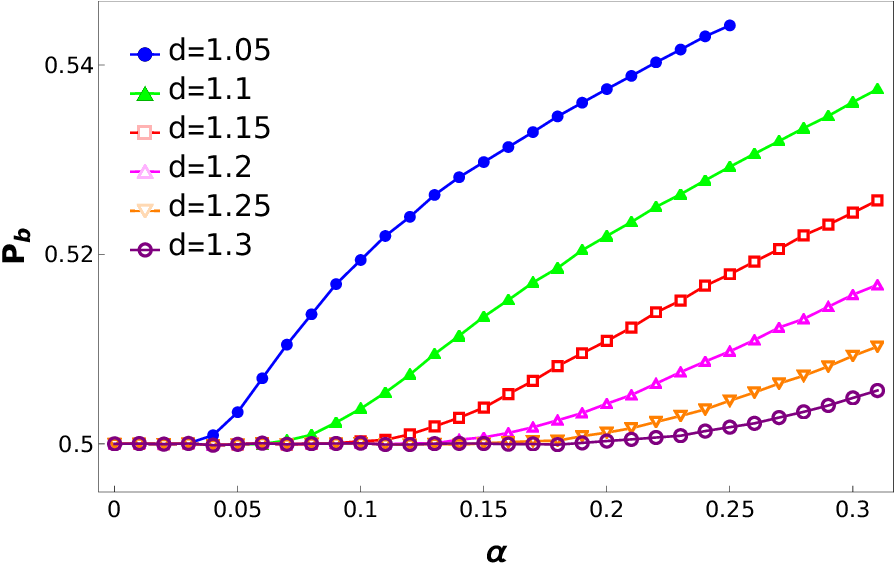}}\hspace{0.3cm}
      \subfigure[]{\includegraphics[scale=0.6]{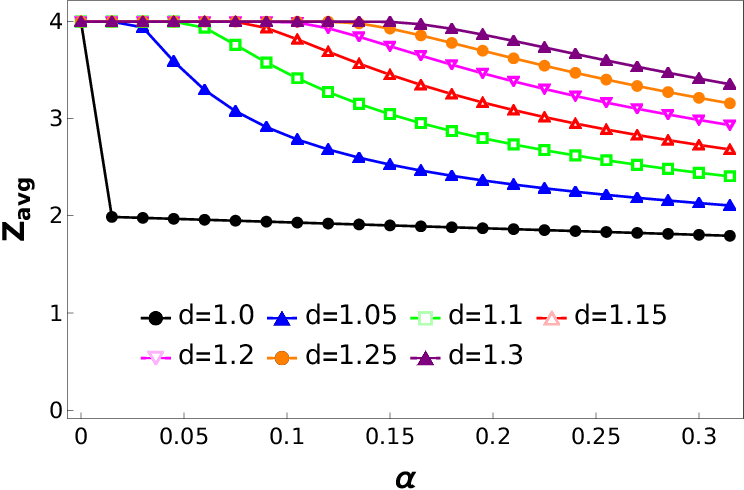}}

\caption{(a) Variation of bond percolation threshold with distortion parameter for distorted square lattice of linear size $L=2^{10}$. Six different curves correspond to different values of $d$. (b) Corresponding variation of the average coordination number $z_\mathrm{avg}(\alpha)$ satisfying $\delta\le d$. Curves for same values of $d$ are represented by same colors in (a) and (b). The black curve for $d=1.0$ in panel (b) shows a sudden jump from $4$ to $2$ for any small $\alpha$. Error bars of the obtained data points are of order $10^{-5}$ and are hidden by the plot markers.The points are joined by lines to aid viewing.}
\label{fig:sap}
\end{figure*}

Since this is a study on bond percolation, all the sites are occupied. On the other hand, a bond can only be occupied if the bond length $\delta$ is less than or equal to the connection threshold $d$. The process of occupying a bond and the determination of the bond percolation threshold $p_\mathrm{b}$ of a finite lattice consist of the following steps:
\begin{enumerate}
\item Begin with a regular lattice of linear size $L$. Identify all the bonds between all the pairs of adjacent sites with a specific numbering scheme.
\item Generate a distorted lattice by dislocating the sites for a certain value of the distortion parameter $\alpha$.
\item Set a value for the connection threshold $d$.
\item Randomly select a bond.
\item If the length of the bond $\delta$ (i.e.\ the distance between the pair of nearest neighbor sites connected by the bond) satisfies $\delta\le d$, occupy the bond. Otherwise, do not occupy and select another bond.
\item After each occupation, put the occupied bond into proper bond-cluster and check whether a spanning cluster exists.
\item If a spanning cluster does not exist, repeat steps 4-6.
\item Stop when a spanning cluster is found and calculate the fraction of occupied bonds $f_\mathrm{b}$.
\item Generate another realization with same $\alpha$ and $d$ and repeat steps 4-8.
\item An average of $1000$ such values of $f_\mathrm{b}$ gives $p_\mathrm{b}(\alpha,d)$
\end{enumerate}

The above process takes away some freedom of selecting the bond for occupation. In regular bond percolation, any unoccupied bond can be occupied randomly. But in this model, only the bonds with bond length $\delta\le d$ can be chosen for occupation. In other words, the number of available empty bonds are reduced due to connection criteria. The main objective of this study is to observe the impact of this restriction on bond percolation threshold. The cluster numbering and spanning analysis have been done by the Newman-Ziff algorithm \cite{Newman1,Newman2}. 

\section{Variation of bond percolation threshold} \label{sec:pb}
In this section we show how the bond percolation threshold varies with distortion parameter $\alpha$ and connection threshold $d$ for distorted square and triangular lattices. The variations are shown in Figs.\ref{fig:sap}, \ref{fig:sdp}, \ref{fig:tap}, and \ref{fig:tdp}. Each data point of these figures represents an average over $1000$ independent realizations of distorted square or triangular lattice of length $L=2^{10}$. Since our focus here is on revealing the variation patterns of $p_\mathrm{b}$, we need to calculate $p_\mathrm{b}(\alpha,d)$ for a large number of combinations of $\alpha$ and $d$. By taking average over $1000$ realizations we get results correct up to fourth decimal place, which is enough for our purposes. For example, the green triangle for $\alpha=0.2$ and $d=1.1$ in Fig.\ref{fig:sap}(a) shows  $p_\mathrm{b}^\mathrm{Sq}(\alpha=0.2,d=1.1)=0.52191(7)$ and the red triangle for $\alpha=0.2$ and $d=1.1$ in Fig.\ref{fig:tap}(a) shows  $p_\mathrm{b}^\mathrm{Tr}(\alpha=0.2,d=1.1)=0.37815(7)$. These values are also close to the corresponding results for infinite lattices given in Table \ref{Tab:precise}.  
\subsection{Distorted Square Lattice}\label{sec:sqpb}
\begin{figure*}
	  \subfigure[]{\includegraphics[scale=0.6]{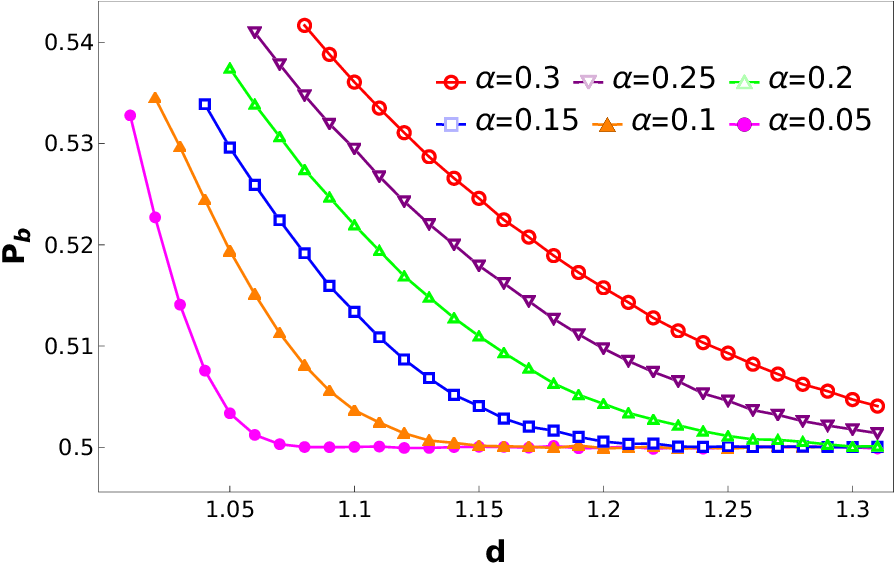}}\hspace{0.3cm}
      \subfigure[]{\includegraphics[scale=0.45]{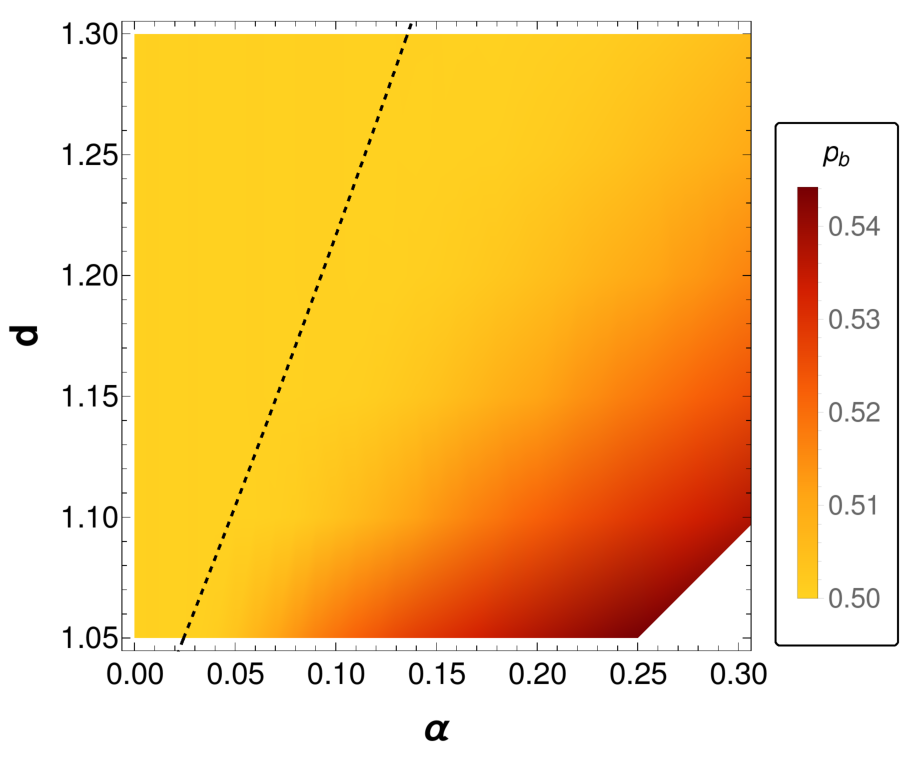}}

\caption{(a) Variation of bond percolation threshold with connection threshold for distorted square lattice of length $L=2^{10}$. Different curves correspond to different values of $\alpha$. Error bars of the obtained data points are of order $10^{-5}$ and are hidden by the plot markers. The points are joined by lines to aid viewing. (b) Variation of bond percolation threshold with distortion and connection threshold for a distorted square lattice. Magnitude of $p_\mathrm{b}(\alpha , d)$ is represented by color variation. White portions indicate regions of no spanning. The dashed line indicates $\delta_\mathrm{Max}^\mathrm{Sq}(\alpha)$ from Eq. \ref{eq:minmax}.}
\label{fig:sdp}
\end{figure*}

As seen in case of site percolation in a distorted square lattice \cite{Sayantan1}, the spanning is not possible for even a slight amount of distortion when the $d\le$ lattice constant ($=1$) for bond percolation also. For $d>1$, the bond percolation threshold $p_\mathrm{b}$ stays at the analytical value $1/2$ for low values of $\alpha$. This is because $d>\delta_\mathrm{Max}^\mathrm{Sq}$, and all the bonds are allowed. As $\alpha$ increases, some bonds cease to be occupied due to increased bond length. Consequently, $p_\mathrm{b}$ starts to increase. This increment starts at a higher value of $\alpha$ as $d$ increases (see Fig.\ref{fig:sap}(a)). Observation of the plots reveals that the spanning becomes difficult with distortion.

To have a deeper insight into the nature of the percolation threshold $p_\mathrm{b}(\alpha)$, we plot the average coordination number $z_\mathrm{avg}(\alpha)$ (Fig.\ref{fig:sap}(b)) of the distorted square lattice. Two nearest neighboring sites remain as neighbors only if $\delta\le d$ is satisfied. It shows that $z_\mathrm{avg}(\alpha)$ is a decreasing function, producing plots of opposite nature to those of $p_\mathrm{b}(\alpha)$. Thus it is clear that the reduction in the average number of neighbors causes difficulty in building a spanning cluster. Fig.\ref{fig:sap}(b) has an extra plot for $d=1.0$ (the black one). There is no corresponding plot Fig.\ref{fig:sap}(a) because spanning is not possible in this case for even a slight distortion. It may be seen that there is a sudden jump of  $z_\mathrm{avg}(\alpha)$ from $4$ to $2$ as $\alpha$ takes a very small non-zero value. It then shows a very slow linear decreasing trend. We note two important observations: (i) when connection threshold $d$ is equal to the lattice constant, $z_\mathrm{avg}(\alpha)$ jumps to half of its original value for a very slight distortion, and (ii) spanning is not possible when $z_\mathrm{avg}\le 2$. We shall come back to this point for triangular lattice.

Variation of $p_\mathrm{b}$ with $d$ is shown in Fig.\ref{fig:sdp}(a). Different curves represent different values of $\alpha$. The reducing nature of each curve demonstrates that spanning becomes easier as $d$ increases. The curves become steeper for lower values of $\alpha$. As expected, all the curves approaches the regular percolation threshold $1/2$ for higher values of the connection threshold $d$. 

The density plot [Fig.\ref{fig:sdp}(b)] for $p_\mathrm{b}(\alpha,d)$ shows the variation of the percolation threshold with both connection threshold and distortion. Higher $p_\mathrm{b}$ is obtained for high $\alpha$ and low $d$. The blank portion on the bottom-right corner indicates that the system cannot percolate for very high distortion and low connection threshold. The dashed line indicates $\delta_\mathrm{Max}^\mathrm{Sq}(\alpha)$. The region to the left of this line indicates $p_\mathrm{b}=1/2$, since $d>\delta_\mathrm{Max}^\mathrm{Sq}$ and all the bonds satisfy the connection criterion.
\subsection{Distorted Triangular Lattice}\label{sec:tripb}

\begin{figure*}
	  \subfigure[]{\includegraphics[scale=0.45]{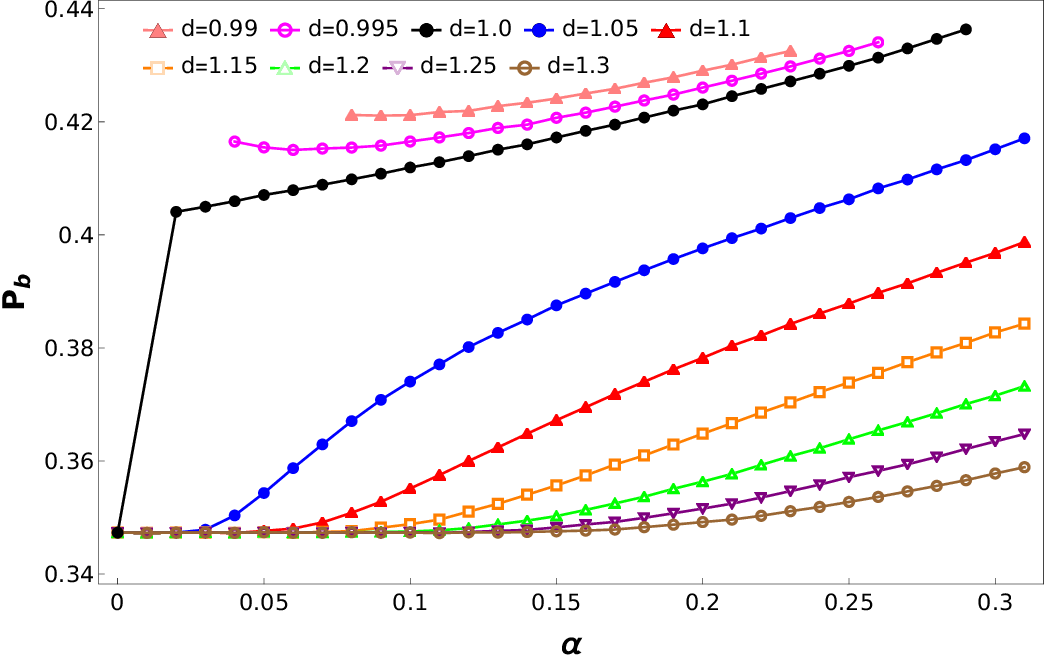}}\hspace{0.3cm}
      \subfigure[]{\includegraphics[scale=0.5]{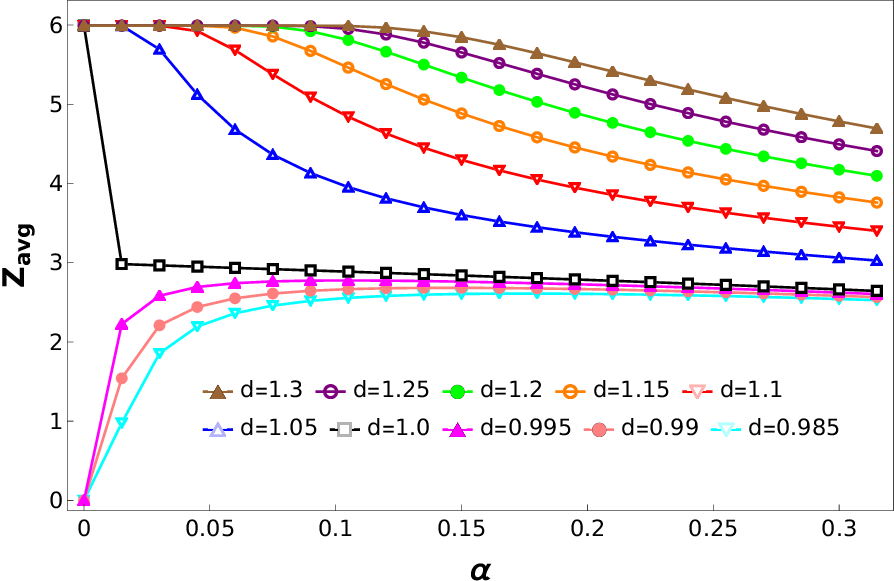}}

        \caption{(a) Variation of bond percolation threshold with distortion parameter for distorted triangular lattice of length $L=2^{10}$. Nine different curves correspond to different values of $d$. (b) Corresponding variation of the average coordination number $z_\mathrm{avg}(\alpha)$ satisfying $\delta\le d$. Curves for same values of $d$ are represented by same colors in (a) and (b). Error bars of the obtained data points are of order $10^{-5}$ and are hidden by the plot markers. The points are joined by lines to aid viewing.}
        \label{fig:tap}
\end{figure*}
Variation patterns of the bond percolation threshold $p_\mathrm{b}$ with distortion in distorted triangular lattice are similar with those for the distorted square lattice when $d>1$. $p_\mathrm{b}$ stays at $2\sin (\pi/18)$ (bond percolation threshold for regular triangular lattice) for low values of $\alpha$. As $\alpha$ increase, $p_\mathrm{b}$ starts to increase as shown in Fig.\ref{fig:tap}(a). Different curves are for different fixed values of $d$. However, there are some striking differences for the triangular case.

\begin{enumerate}
\item We {\it do} get spanning for $d\le 1$ for this lattice.
\item For $d=1$, $p_\mathrm{b}(\alpha)$ jumps discontinuously from $2\sin (\pi/18)$ at $\alpha=0$ to a significantly higher value at any small value of $\alpha$. The $p_\mathrm{b}(\alpha)$ curve cannot be made continuous at $\alpha=0$ by reducing $\alpha$ to even smaller values.
\item For $d<1$, $p_\mathrm{b}$ reduces with distortion at low-$\alpha$ region, forms a minimum, then starts to increase as $\alpha$ increase. This is most clearly seen for the $d=0.995$ curve. This means, when the connection threshold is less than the lattice constant, a low amount of distortion in fact could help spanning.
\end{enumerate}

\begin{figure*}
	  \subfigure[]{\includegraphics[scale=0.6]{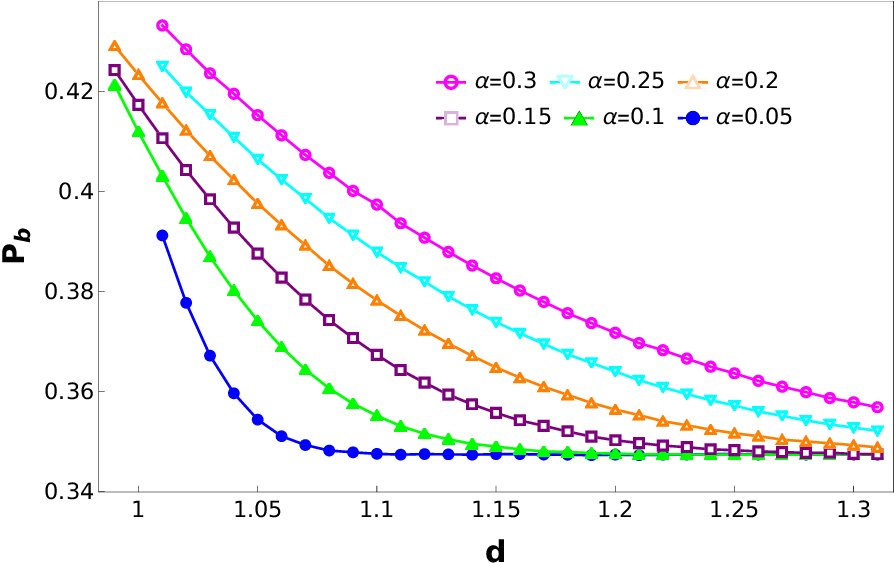}}\hspace{0.3cm}
      \subfigure[]{\includegraphics[scale=0.45]{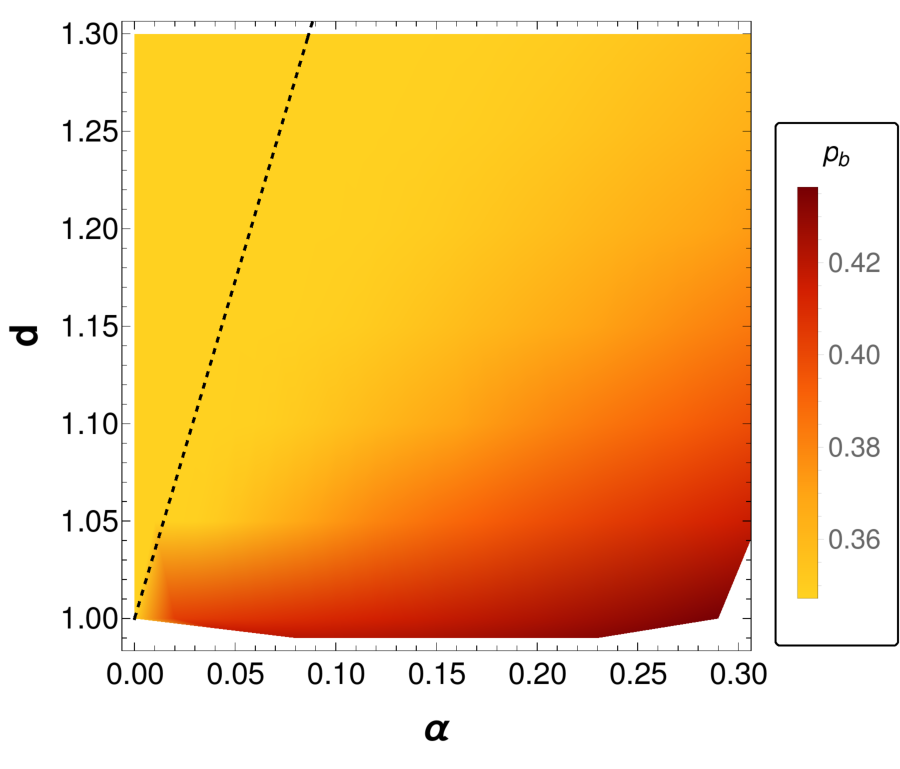}}

\caption{(a) Variation of bond percolation threshold with connection threshold for distorted triangular lattice of length $L=2^{10}$. Error bars of the obtained data points are of order $10^{-5}$ and are hidden by the plot markers. The points are joined by lines to aid viewing. (b) Variation of bond percolation threshold with distortion and connection threshold for a distorted triangular lattice. Magnitude of $p_\mathrm{b}(\alpha , d)$ is represented by color variation. White portions indicate regions of no spanning. The dashed line indicates $\delta_\mathrm{Max}^\mathrm{Tr}(\alpha)$ from Eq. \ref{eq:minmax}.}
\label{fig:tdp}
\end{figure*}

All the above findings may be understood from the plots of $z_\mathrm{avg}(\alpha)$ (satisfying $\delta\le d$) shown in Fig.\ref{fig:tap}(b). For $d>1.0$, the growth of $p_\mathrm{b}(\alpha)$ is explained by the reduction of $z_\mathrm{avg}(\alpha)$. For $d=1.0$, the discontinuous rise of $p_\mathrm{b}(\alpha)$ corresponds to the discontinuous drop of $z_\mathrm{avg}(\alpha)$. In this case, the jump is from $6$ to $3$. So, for both square and triangular lattices, $z_\mathrm{avg}(\alpha)$ drops to the half of its original value. However, this time $z_\mathrm{avg}(\alpha)$ stays close to 3, which allows spanning. The plots for $d<1.0$ show an initial increasing pattern explaining the reverse nature of $p_\mathrm{b}(\alpha)$. The increment in average nearest neighbors is caused by the reduction of $\delta_\mathrm{min}^\mathrm{Tr}$ for low $\alpha$ (see Eq. \ref{eq:minmax}), which allows more bonds to satisfy the connection criterion. Again note that, spanning is only possible when $z_\mathrm{avg}(\alpha)$ stays well above 2.

The variation of $p_\mathrm{b}$ with $d$ for different fixed values of $\alpha$ is shown in Fig.\ref{fig:tdp}(a). Not surprisingly, $p_\mathrm{b}$ reduces as $d$ is increased. Crossing at very small values of $d$ is a manifestation of the fact discussed in point no. 3 above.

A similar density plot [Fig.\ref{fig:tdp}(b)] for $p_\mathrm{b}(\alpha,d)$ is shown for the distorted triangular lattice. The blank portions at the bottom-left and bottom-right corners indicate regions of no spanning. The dashed line indicates $\delta_\mathrm{Max}^\mathrm{Tr}(\alpha)$. The region to the left of this line indicates $p_\mathrm{b}=2\sin (\pi /18)$, the bond percolation threshold for the regular triangular lattice.
\subsection{Precise thresholds}\label{sec:pbprecise}

\begin{figure*}
      \subfigure[]{\includegraphics[scale=0.59]{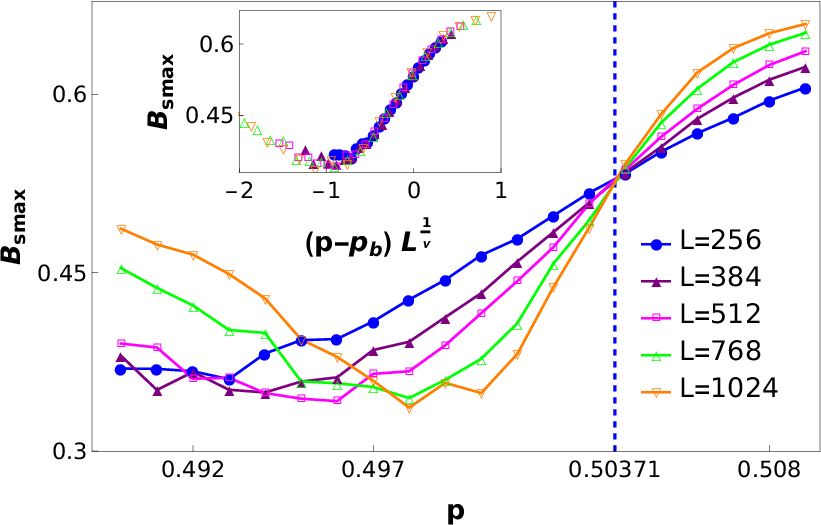}}
      \subfigure[]{\includegraphics[scale=0.58]{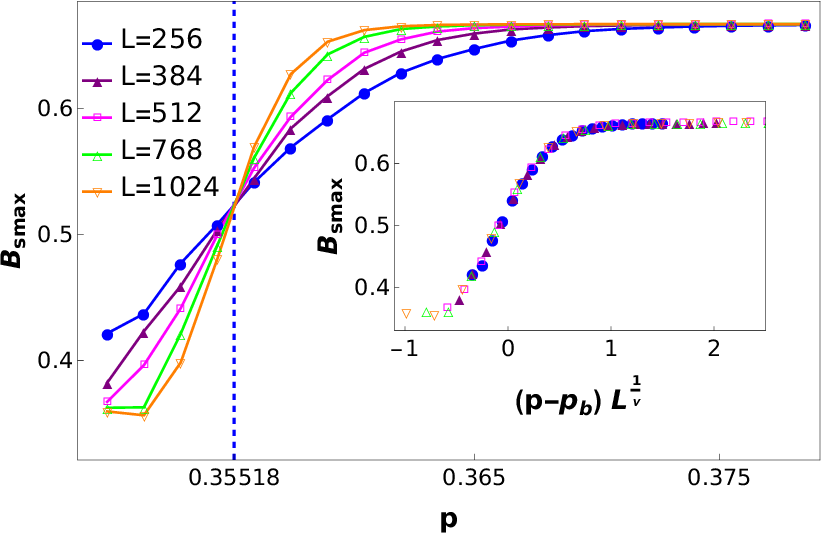}}
\caption{Plots of Binder cumulant for (a) distorted square and (b) distorted triangular lattices. The values for the distortion parameter and the connection threshold are $\alpha=0.1$ and $d=1.1$ for both the plots. The dotted vertical lines indicate the corresponding values of $p_\mathrm{b}^{\infty}(\alpha,d)$ from Table \ref{Tab:precise} . Error bars of the obtained data points around $p_\mathrm{b}^{\infty}(\alpha,d)$ are of order $5\times 10^{-3}$ and are hidden by the plot markers. The points are joined by lines to aid viewing. The data collapse for Binder cumulant is shown in the insets of (a) and (b) when the horizontal axis is scaled as $(p-p_\mathrm{b})L^{1/\nu}$.}
\label{fig:precise}
\end{figure*}
In this section, more precise values of the bond percolation threshold are determined. The values have been calculated by the intersections of the Binder cumulant curves. 
The Binder cumulant may be defined as
\begin{equation}
B_\mathrm{smax}(p,L)=1-\frac{\langle (S_\mathrm{max}(p,L))^4\rangle}{3\langle (S_\mathrm{max}(p,L))^2\rangle^2},
\end{equation}
where, $S_\mathrm{max}(p,L)$ is the size of the largest cluster in a lattice of length $L$ at occupation probability $p$. Here $\langle\rangle$ denotes the average over many independent realizations of the lattice. In our calculations, we took average over $10^4$ realizations. An average of the intersecting points of the curves of $B_\mathrm{smax}(p)$ for different lattice sizes gives an estimate of percolation threshold of the infinite lattice. This is demonstrated in Fig.\ref{fig:precise}(a) and \ref{fig:precise}(b). The values of the $p_\mathrm{b}^{\infty}(\alpha,d)$ are given in Table \ref{Tab:precise}. These are very close to the corresponding values of Fig.\ref{fig:sap}(a) and \ref{fig:tap}(a) calculated earlier for finite lattices of linear size $L=2^{10}$.

\begin{figure*}
	  \subfigure[]{\includegraphics[scale=0.58]{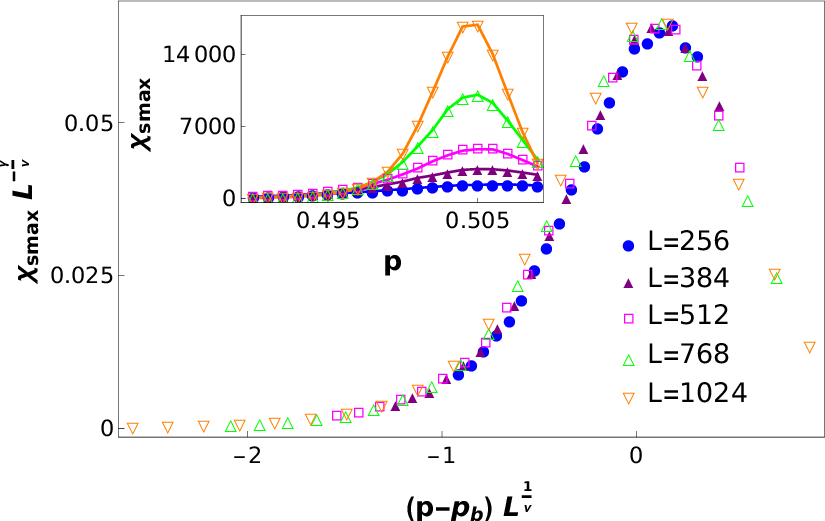}}
      \subfigure[]{\includegraphics[scale=0.58]{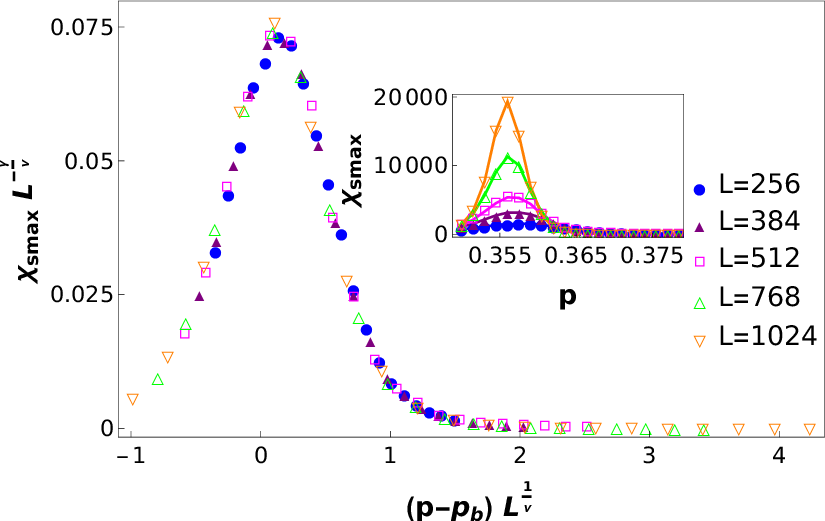}}
      \subfigure[]{\includegraphics[scale=0.58]{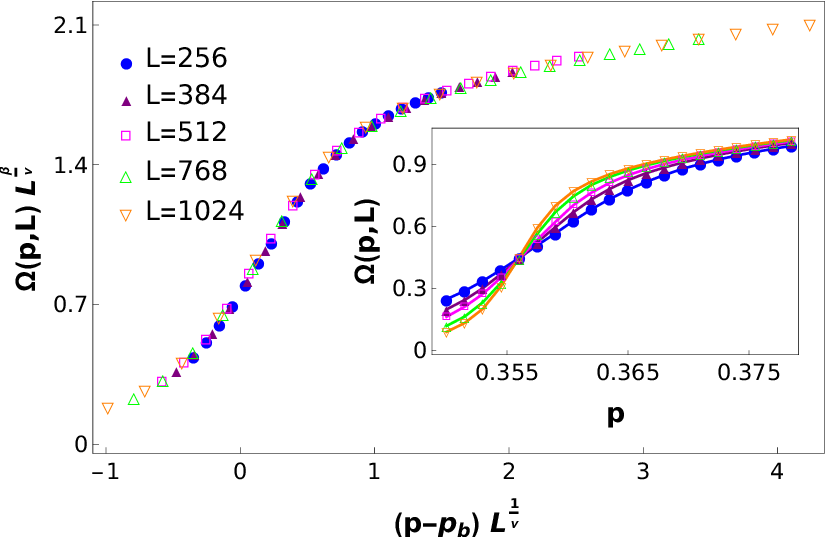}}
      \subfigure[]{\includegraphics[scale=0.58]{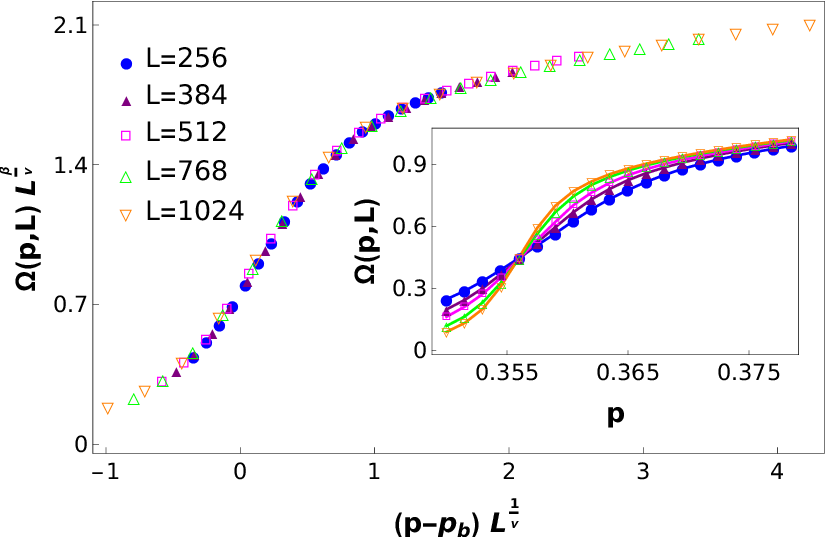}}
\caption{Data collapse of (a) $\chi_\mathrm{smax}(p,L)$ for distorted square and (b) $\chi_\mathrm{smax}(p,L)$ for distorted triangular lattice, (c) $\Omega(p,L)$ for distorted square lattice, and (d) $\Omega(p,L)$ for distorted triangular lattice of different system sizes. The horizontal axis scaled as $(p-p_\mathrm{b})L^{1/\nu}$ for all the plots. The vertical axis of (a) and (b) is scaled as $\chi_\mathrm{smax}L^{-\gamma/\nu}$, whereas the vertical axis for (c) and (d) is scaled as $\Omega L^{\beta/\nu}$. All the plots are generated with $\alpha=0.1$ and $d=1.1$. Error bars of the obtained data points are of order $10^{-3}$ and are hidden by the plot markers. (Insets) Corresponding plots of $\chi_\mathrm{smax}(p,L)$ and $\Omega(p,L)$.}
 \label{collapse}
 \end{figure*}

\begin{table}

\begin{tabular}{|c|c|c|c|c|}
\hline 
\makecell{Lattice\\Type} & $d$ & $\alpha$
& \makecell{Code of Fig.\ref{fig:all_collapse}} & \makecell{$p_\mathrm{b}^{\infty}(\alpha,d)$}  \\ 
\hline 
\makecell{Distorted\\Square} & \makecell{1.1\\1.1\\1.2\\1.2} & \makecell{0.1\\0.2\\0.1\\0.2} & \makecell{$\mathrm{S^{(1)}}$\\$\mathrm{S^{(2)}}$\\$\mathrm{S^{(3)}}$\\$\mathrm{S^{(4)}}$} & \makecell{0.50371(7)\\0.52196(4)\\0.50016(4)\\0.50439(5)}\\ 
\hline 
\makecell{Distorted\\Triangular} & \makecell{1.1\\1.1\\0.995\\0.995} & \makecell{0.1\\0.2\\0.1\\0.2} & \makecell{$\mathrm{T^{(1)}}$\\$\mathrm{T^{(2)}}$\\$\mathrm{T^{(3)}}$\\$\mathrm{T^{(4)}}$}& \makecell{0.35518(4)\\0.37837(7)\\0.41657(9)\\0.42625(6)} \\ 
\hline 
\end{tabular} 

\caption{Precise values of $p_\mathrm{b}(\alpha,d)$ obtained from Binder cumulant. The fourth column indicates the combination codes used in Fig.\ref{fig:all_collapse} . For example, $\mathrm{T^{(2)}}$ refers to $d=1.1$ and $\alpha=0.2$ for distorted triangular lattice.}
\label{Tab:precise}
\end{table}

\section{Universality}\label{sec:univ}
As found for site percolation, it is expected that the bond percolation transition for the distorted lattices should fall into the same universality class as the regular percolation. Nevertheless, it is worthwhile to verify this fact. We use the following standard values of the critical exponents for percolation in two dimensional systems, $\nu=4/3, \gamma=43/18,$ and $\beta=5/36$. In the inset Fig.\ref{fig:precise}(c) and \ref{fig:precise}(d), the data collapse of Binder cumulant has been demonstrated for distorted square and triangular lattices respectively when the horizontal axis is scaled as $(p-p_\mathrm{b})L^{1/\nu}$. This establishes that the value of $\nu$ is unaltered.

In Figs.\ref{collapse}(a) and \ref{collapse}(b), the universality is established through the data collapse of susceptibility, which is defined as 
\begin{equation}
\chi_\mathrm{smax}(p,L)=\langle (S_\mathrm{max}(p,L))^2\rangle-\langle S_\mathrm{max}(p,L)\rangle^2.
\end{equation}
Plots of $\chi_\mathrm{smax}(p)$ for different system sizes are shown in the insets. In this case, the different curves merge when the horizontal and vertical axes are scaled as $(p-p_\mathrm{b})L^{1/\nu}$ and $\chi_\mathrm{smax}L^{-\gamma/\nu}$ respectively. Excellent collapse has been obtained for both distorted square (Fig.\ref{collapse}(a)) and triangular (Fig.\ref{collapse}(b)) lattices. Since we have already seen that $\nu$ is unaltered, we may conclude that $\gamma$ also remains the same.

Figs. \ref{collapse}(c) and \ref{collapse}(d) show data collapse for the order parameter defined as
\begin{equation}
\Omega(p,L)=\frac{\langle S_\mathrm{max}\rangle}{L^2},
\end{equation}
where, $\langle S_\mathrm{max}\rangle$ denotes the configurational average of the largest cluster size. Nice data collapse is obtained for both lattices when horizontal axis and vertical axis are scaled as $(p-p_\mathrm{b})L^{1/\nu}$ and $\Omega L^{\beta/\nu}$ respectively. This demonstrates that $\beta$ does not vary. On the basis of the constancy of $\nu , \gamma$, and $\beta$, we conclude that the bond percolation in distorted two dimensional lattices falls into the universality class of regular percolation.

\begin{figure*}
	  \subfigure[]{\includegraphics[scale=0.35]{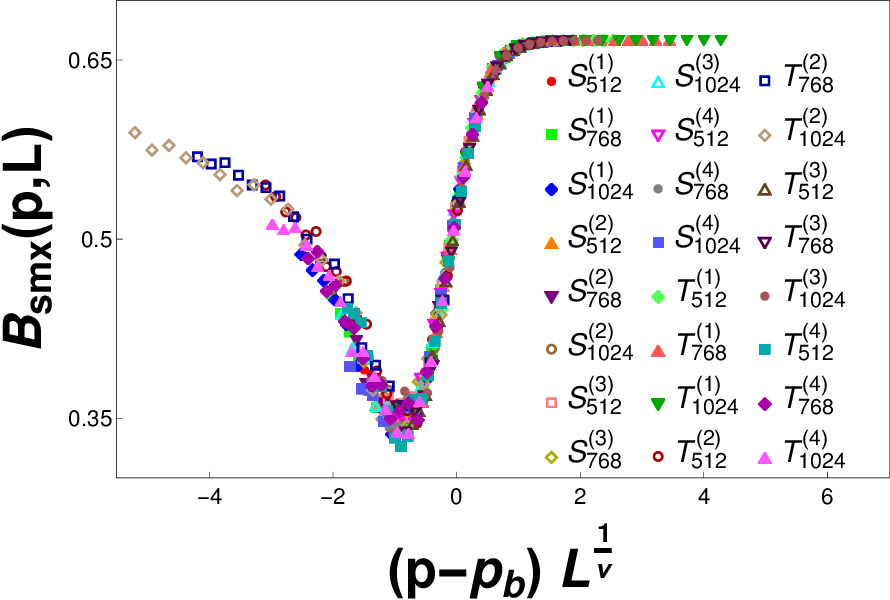}}
      \subfigure[]{\includegraphics[scale=0.35]{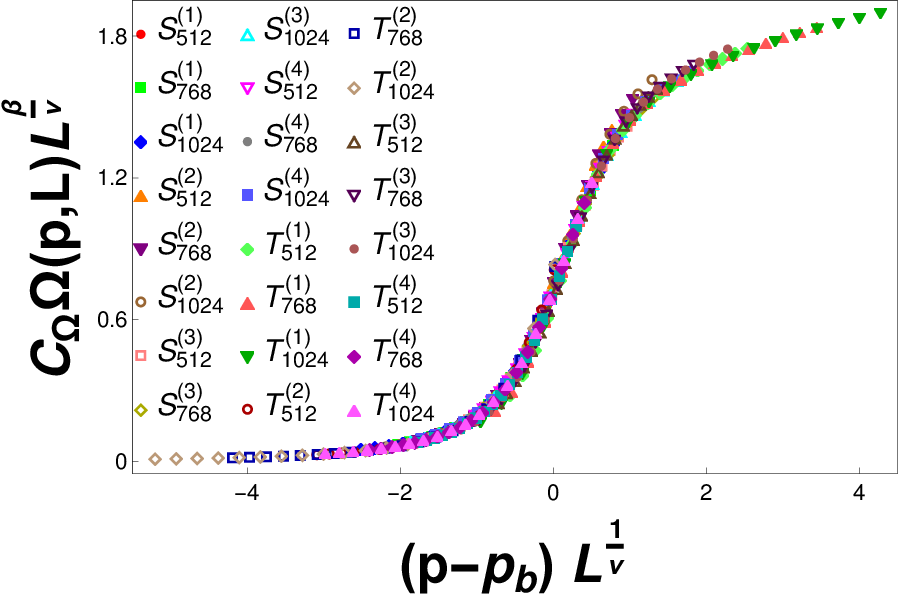}}
      \subfigure[]{\includegraphics[scale=0.35]{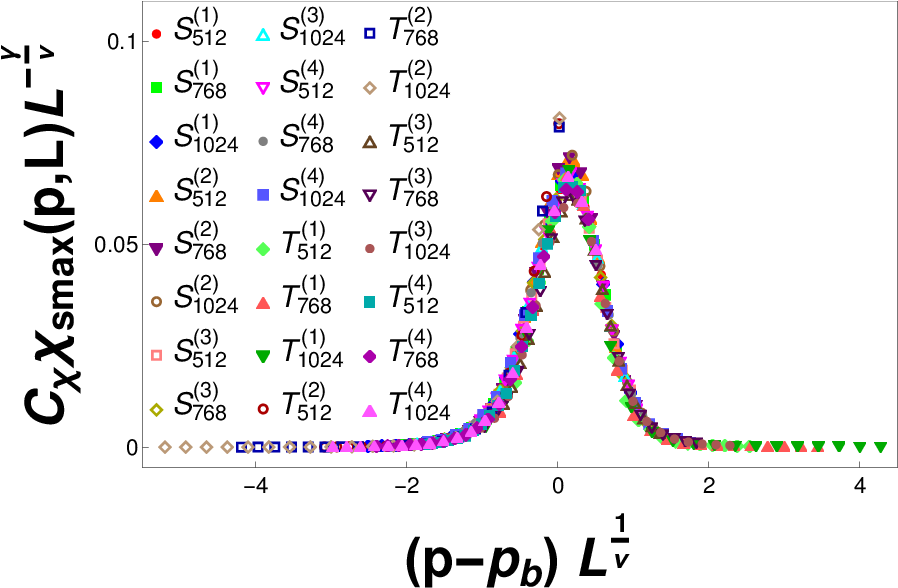}}
\caption{Data collapse of (a) Binder cumulant (b) order parameter and (c) susceptibility demonstrating universality for both the lattices and different combinations of $d$ and $\alpha$. The combination codes are written as $\mathrm{S}^\mathrm{(i)}_L$ and $\mathrm{T}^\mathrm{(i)}_L$ for distorted square and triangular lattice respectively, as given in Table \ref{Tab:precise}. Here, $L$ (=512, 768, 1024) refers to the linear size of the lattices.}
\label{fig:all_collapse}
\end{figure*}

Since the universality of the exponents do not depend on the lattice type, it should  also be possible to obtain data collapse for both lattices together. In Fig. \ref{fig:all_collapse}, data collapse is shown for (a) $B_\mathrm{smax}(p,L)$, (b) $\Omega(p,L)$, and (c) $\chi_\mathrm{smax}(p,L)$. In each plot, three lattice sizes ($L=$512, 768, 1024) of eight different lattice-$\alpha$-$d$ combinations of Table \ref{Tab:precise} are considered. So, each sub-figure of Fig.\ref{fig:all_collapse} shows collapse of 24 sets of data. 

Fig. \ref{fig:all_collapse}(a) shows nice collapse for Binder cumulant. The scenario is slightly different for order parameter and susceptibility. While the six combinations with $d>1$ (see Table \ref{Tab:precise}) collapse with each other, the two combinations with $d<1$ for distorted triangular lattice, namely, $\mathrm{T^{(3)}}$ and $\mathrm{T^{(4)}}$, require rescaling of the vertical axis by a factor. We call this factor as $C_{\Omega}$ and $C_{\chi}$ corresponding to $\Omega(p,L)$ (Fig.\ref{fig:all_collapse}(b)) and $\chi_\mathrm{smax}(p,L)$ (Fig. \ref{fig:all_collapse}(c)) respectively. We find nice collapse for $C_{\Omega}=0.6$ for $\mathrm{T^{(3)}}$ and 0.8 for $\mathrm{T^{(4)}}$. Similarly, good collapse was obtained for $C_{\chi}$= 0.4 and 0.6 for $\mathrm{T^{(3)}}$ and $\mathrm{T^{(4)}}$ respectively.
\section{The critical connection threshold}\label{sec:dc}
\begin{figure*}
	  \subfigure[]{\includegraphics[scale=0.4]{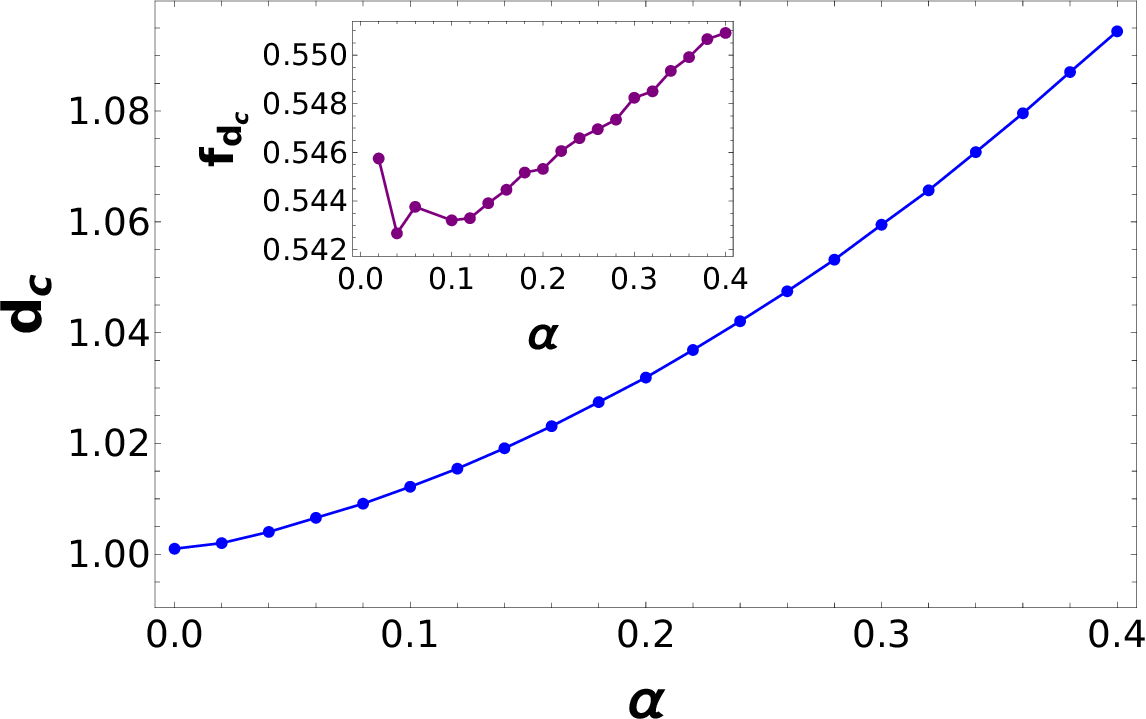}}\hspace{0.3cm}
      \subfigure[]{\includegraphics[scale=0.4]{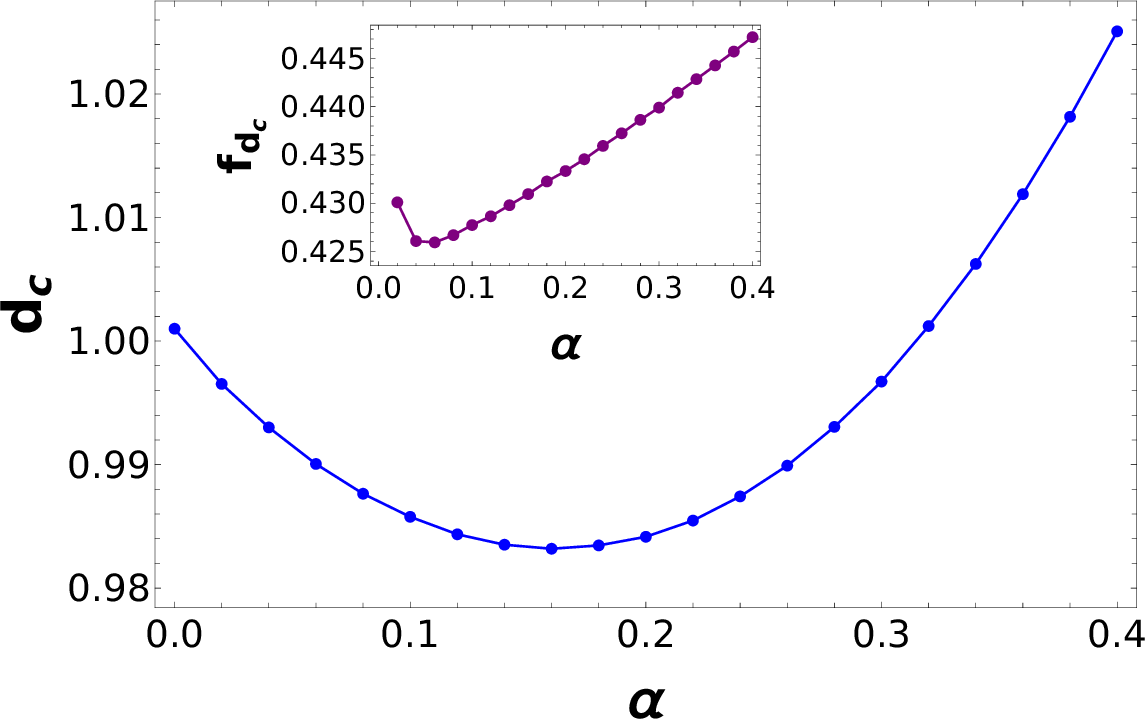}}
\caption{Variation of critical connection threshold with distortion parameter for distorted (a) square and (b) triangular lattices. (Insets) Fraction of the occupied bonds $f_{d_\mathrm{c}}(\alpha)$ satisfying $\delta\le d_\mathrm{c}$. Error bars of the obtained data points are of order $10^{-5}$ and are hidden by the plot markers. The points are joined by lines to aid viewing.}
\label{fig:adc}
\end{figure*}
We now introduce a quantity what may be called the critical connection threshold $d_\mathrm{c}$. We have already found that no spanning cluster exists in a distorted square lattice when the connection threshold $d\le$ lattice constant (which we considered to be $1$). This is a striking result. According to the connection criterion of our model, all the bonds of a regular lattice are occupied when $d=1$ because the adjacent sites are separated by a distance equal to the lattice constant($=1$). Our results show that if a square lattice is even very slightly distorted (keeping $d=1$), spanning is totally disrupted. To achieve spanning in such a lattice, one must set $d>1$ allowing connectivity of the longer bonds. The question is, what should be the minimum allowed bond distance to achieve spanning? We call this distance as the critical connection threshold $d_\mathrm{c}$. Of course, this must be a function of the distortion parameter $\alpha$.

$d_\mathrm{c}$ is numerically estimated in the following way. 
\begin{enumerate}
\item To begin with, a distorted lattice is formed with a fixed value of $\alpha$ and $d$.
\item {\it All} the bonds satisfying $\delta \le d$ are occupied.
\item Initially, the connection threshold $d$ is kept at a low value. The value is so low that spanning is not possible even though the condition 2 is applied.
\item $d$ is now slowly increased. Spanning is checked for each value of $d$.
\item The first value of $d$, when spanning is obtained, is noted.
\item Hence a new configuration is generated for the same value of $\alpha$ and whole process is repeated.
\item An average of $1000$ such values of $d$ is identified as the critical connection threshold $d_\mathrm{c}$.
\end{enumerate}

In Fig.\ref{fig:adc}(a) we plot $d_\mathrm{c}(\alpha)$ for the distorted square lattice, which shows a monotonic growth. As can be noticed from the figure, $d_\mathrm{c}$ is slightly higher than $1$ at $\alpha=0$. This is expected as we have already seen that no spanning is possible in a regular lattice when the connection threshold $\le$ lattice constant. $d_\mathrm{c}(\alpha)$ then grows steadily in a parabolic fashion. Therefore, we conclude that the minimum allowed bond distance must increase with distortion.

The plot of the critical connection threshold $d_\mathrm{c}$ with distortion for the triangular lattice (Fig.\ref{fig:adc}(b)) is markedly different from that for the square lattice (Fig.\ref{fig:adc}(a)). It is seen that $d_\mathrm{c}$ initially decreases, forms a minimum, then increases as $\alpha$ increases. The minimum value of $d_\mathrm{c}$ is approximately $0.985$ at $\alpha\sim 0.16$. The underlying reason is explained below. For small values of $\alpha$, the minimum displacement in the triangular lattice, denoted as $\delta_\mathrm{min}^\mathrm{Tr}$ starts at 1.0 and decreases steadily with increasing $\alpha$ (see Eq. \ref{eq:minmax}). As may be seen in Fig.\ref{fig:tap}(b), this reduction in $\delta_\mathrm{min}^\mathrm{Tr}$ increases the average number of nearest neighbors $z_\mathrm{avg}(\alpha)$ for $d<1$, facilitating the formation of a spanning cluster. Consequently, the critical connection threshold $d_\mathrm{c}$ decreases with $\alpha$ in the low-$\alpha$ regime. This reduction is absent in square lattice since there is no spanning region for $d<1.0$. On the other hand, for larger values of $\alpha$, $z_\mathrm{avg}(\alpha)$ decreases steadily with $\alpha$. This gives rise to an increasing trend in $d_\mathrm{c}(\alpha)$ for $\alpha>0.16$.

To investigate further, we measure the fraction of occupied bonds $f_{d_\mathrm{c}}$ satisfying $\delta\le d_\mathrm{c}$, for both the lattices. To calculate this, we first determine the value of $d_\mathrm{c}$ at a certain value of $\alpha$. Hence we generate a distorted lattice with that value of $\alpha$ and calculate the fraction of bonds with bond length $\delta\le d_\mathrm{c}$. Averaging over $1000$ such configurations we get $f_{d_\mathrm{c}}(\alpha)$. The plots are shown in the insets of Fig.\ref{fig:adc}. We notice the following interesting facts from these plots:
\begin{enumerate}
\item The fraction of bonds $f_{d_\mathrm{c}}$ for both the lattices are different from the values $0.5$ and $2\sin (\pi /18)$ for regular square and triangular lattices respectively. Therefore, it may be concluded that the bond percolation in distorted lattices is an independent and unique model, likely due to the correlations in the occupied bonds -- they are not independent.
\item $f_{d_\mathrm{c}}$ is not constant. After an initial decline it increases very slowly but steadily for both the lattices. The whole range of values are significantly higher than the regular percolation thresholds.
\end{enumerate}
\section{Discussion in the Perspective of Prior Research}\label{sec:perspective}
To place the present results in a broader context, it is useful to compare them with earlier studies on percolation in distorted lattices. While the current work addresses bond percolation in distorted square and triangular lattices, our previous investigations \cite{Sayantan1,Sayantan2, Sayantan3} focused on site percolation in square and simple cubic lattices. Among these, only Refs. \cite{Sayantan1} and \cite{Sayantan2} are directly comparable, since Ref. \cite{Sayantan3} considered link formation with a flexible number of neighbors, which lies beyond the scope of this article. The key points of comparison are as follows:
\begin{enumerate}
\item Methodological aspects: The procedure for generating distorted lattices is identical to that used in our previous studies. However, the percolation mechanisms differ fundamentally between site and bond cases. In regular site (bond) percolation, all bonds (sites) are assumed preoccupied, and clusters are formed by adjacent occupied sites (bonds). In distorted lattices, sites are shifted from their regular positions, and a bond between two sites can only be occupied if their separation $\delta$ is less than the connection threshold $d$. Thus, in distorted site percolation, all sites are eligible for occupation while bonds satisfying $\delta<d$ are preoccupied; whereas in distorted bond percolation, all sites are preoccupied and only bonds satisfying $\delta<d$ are eligible for occupation. Consequently, the algorithms for neighborhood management and cluster building are substantially different in the two cases.
\item Distorted square lattice: A direct comparison between site \cite{Sayantan1} and bond (present work) percolation can be made for distorted square lattices. In both cases, spanning becomes increasingly difficult with greater distortion, as reflected in the monotonic rise of the percolation thresholds $p_c(\alpha)$ and $p_b(\alpha)$. However, the magnitude of threshold shifts differs: in bond percolation the sensitivity to distortion appears weaker. A notable common feature is that no spanning cluster is observed, even under the slightest distortion, when the connection threshold satisfies $d\le 1$. In contrast, spanning is still possible for $d\le 1$ in distorted triangular lattices, similar to the behavior observed earlier for site percolation in distorted simple cubic lattices \cite{Sayantan2}.
\item Universality: In earlier studies we showed that site percolation in distorted systems falls within the universality class of regular percolation. The present results extend this conclusion to bond percolation in two-dimensional distorted lattices. The critical exponents $\beta$, $\nu$, and $\gamma$ retain their universal values, and the demonstration here is more robust: data collapse is obtained for Binder cumulant, susceptibility, and order parameter across different system sizes for both lattices, separately and jointly.
\item Critical connection threshold: A new quantity introduced in this work is the critical connection threshold $d_c$, defined as the minimum bond length required for spanning. Interestingly, the functional dependence of $d_c(\alpha)$ differs significantly between square and triangular lattices, providing further insight into the interplay between lattice geometry and distortion.
\item General trends: Taken together, the four studies on distorted lattices demonstrate that both site and bond percolation thresholds are strongly affected by lattice distortion. While the qualitative features for square lattices suggest a weaker impact in the bond percolation case, a combination of similarities and distinct behaviors is observed across different lattices and percolation types. Importantly, all cases reaffirm that percolation in distorted systems continues to belong to the universality class of regular percolation.
\end{enumerate}
\section{Summary}\label{sec:summary}
To summarize, we have studied bond percolation properties of distorted square and triangular lattices. The distortion is achieved by random but tunable dislocations of the sites from their regular lattice positions through the distortion parameter $\alpha$. The bonds, whose lengths thus become greater than the connection threshold $d$, cannot be occupied. This restriction, along with the interplay between $\alpha$ and $d$ give rise to a number of interesting results summarized below.
\begin{enumerate}
\item When $d$ is greater than the lattice constant (set to be 1), the percolation threshold $p_\mathrm{b}$ increases with distortion for both the lattices.
\item If $d\le 1$, no spanning cluster can be found for the square lattice. Whereas, the plots of $p_\mathrm{b}(\alpha)$ for triangular lattice show a discontinuous drop for $d=1$ and a decreasing trend for $d<1$ in the low-$\alpha$ regime.
\item All these behaviors can be attributed to the variation of average coordination number $z_\mathrm{avg}(\alpha,d)$ of the distorted lattice.
\item Bond percolation threshold decreases with $d$ for both the lattices. 
\item Precise values of $p_\mathrm{b}^{\infty}(\alpha,d)$ obtained through the Binder cumulant are close to $p_\mathrm{b}^{1024}(\alpha,d)$ for both the lattices.
\item The transition falls into the same universality class of the regular percolation. Universality has been demonstrated by obtaining nice data collapse for Binder cumulant, order parameter and susceptibility with different $\alpha$-$d$ combinations for both the lattices together.
\item The critical connection threshold $d_\mathrm{c}$ is defined as the minimum value of the connection threshold $d$ required to obtain spanning when all the bonds satisfying the connection criterion $\delta\le d$ are occupied. Plots of $d_\mathrm{c}(\alpha)$ are markedly different for the two lattices. 
\item The results of this study should be applicable to many natural systems and real lattices since they are hardly found in a perfectly ordered state. As an example, one may consider forest fire mitigation strategies for planted forests. A distorted array may significantly reduce fire propagation as indicated in Figs. \ref{fig:sap} and \ref{fig:tap}. This idea was discussed in the context of site percolation \cite{Sayantan1}. Bond percolation becomes relevant when the focus is on the connections between the burning trees.
\item The ideas and the techniques of this work may be extended to other systems in the future. For example, an immediate extension may be to study bond percolation in distorted simple cubic, face centered cubic, and body centered cubic lattices. We believe that these three dimensional systems would yield more rich and complex behavior. 
\end{enumerate}

\begin{acknowledgments}
This work is funded by Anusandhan National Research Foundation (ANRF), Department of Science and Technology, Government of India. The project file No. is SUR/2022/002345. SM gratefully acknowledges financial support through a National Post Doctoral Fellowship from  ANRF under  project file no. PDF/2023/002952. The authors thank Dipa Saha and Abdur Rashid Miah for useful discussion. The computation facilities availed at the Department of Physics, University of Gour Banga, Malda are gratefully acknowledged.
\end{acknowledgments}

\end{document}